\documentclass[twocolumn,floatfix,aps,superscriptaddress, nofootinbib]{revtex4-2}
\usepackage{graphicx}
\usepackage{amsmath}
\usepackage{amssymb}
\usepackage{bm}
\usepackage{hyperref}

\usepackage{color}

\begin{document}

\title{Helical Luttinger liquid on a space-time lattice}
\author{V. A. Zakharov}
\affiliation{Instituut-Lorentz, Universiteit Leiden, P.O. Box 9506, 2300 RA Leiden, The Netherlands}
\author{J. Tworzyd{\l}o}
\affiliation{Faculty of Physics, University of Warsaw, ul.\ Pasteura 5, 02--093 Warszawa, Poland}
\author{C. W. J. Beenakker}
\affiliation{Instituut-Lorentz, Universiteit Leiden, P.O. Box 9506, 2300 RA Leiden, The Netherlands}
\author{M. J. Pacholski}
\affiliation{Max Planck Institute for the Physics of Complex Systems, N\"{o}thnitzer Strasse 38, 01187 Dresden, Germany}

\date{January 2024}

\begin{abstract}
The Luttinger model is a paradigm for the breakdown due to interactions of the Fermi liquid description of one-dimensional massless Dirac fermions. Attempts to discretize the model on a one-dimensional lattice have failed to reproduce the established bosonization results, because of the fermion-doubling obstruction: A local and symmetry-preserving discretization of the Hamiltonian introduces a spurious second species of low-energy excitations, while a nonlocal discretization opens a single-particle gap at the Dirac point. Here we show how to work around this obstruction, by discretizing both space and time to obtain a \textit{local} Lagrangian for a helical Luttinger liquid with Hubbard interaction. The approach enables quantum Monte Carlo simulations that preserve the topological protection of an unpaired Dirac cone.
\end{abstract}
\maketitle

\begin{figure}[tb]
\centerline{\includegraphics[width=0.8\linewidth]{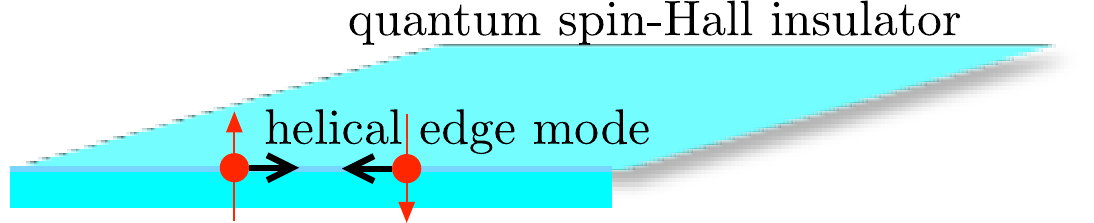}}
\caption{Helical edge mode, consisting of counter-propagating spin-up and spin-down electrons on the 1D boundary of a 2D quantum spin Hall insulator.
}
\label{fig_layout}
\end{figure}

{\em Introduction ---}
A quantum spin Hall insulator \cite{Mac11} supports a one-dimensional (1D) helical edge mode of counterpropagating massless electrons (Dirac fermions, see Fig.\ \ref{fig_layout}), with a linear dispersion $E=\pm\hbar v_{\rm F}k$. The crossing at momentum $k=0$ (the Dirac point) is protected from gap-opening \cite{Kan05} --- provided that there is only a single species of low-energy excitations and provided that fundamental symmetries (time-reversal symmetry, chiral symmetry) are preserved. This topological protection is broken on a lattice by fermion doubling \cite{Tong}: Any local and symmetry-preserving discretization of the momentum operator $k=-i\hbar d/dx$ must introduce a spurious second Dirac point \cite{Nie81,Che17}.

Fermion doubling is problematic if one wishes to study interaction effects of 1D massless electrons (a Luttinger liquid \cite{Lut63,Hal81,Gia03,Giu08}) by means of a lattice fermion method such as quantum Monte Carlo \cite{Sca81,Hir83,Hir86,Gub16,Bec17}. A way to preserve the time-reversal and chiral symmetries on a lattice is to increase the dimensionality of the system \cite{Kap24a,Kap24b}. One can simulate a 2D system in a ribbon geometry, so that the two fermion species are spatially separated on opposite edges \cite{Hoh11,Li11,Zhe11,Yam11,Kim15}. The 2D simulation is computationally more expensive than a fully 1D simulation, but more fundamentally, the presence of states in the bulk may obscure the intrinsically 1D physics of a Luttinger liquid \cite{Hoh12}. A 1D simulation using a nonlocal spatial discretization \cite{Dre76} that avoids fermion doubling was studied recently \cite{Wan23}, without success: The nonlocality gaps the Dirac point \cite{Wan23,Lia23}.

Here we show that it can be done: A 1D helical Luttinger liquid can be simulated on a lattice if both space \textit{and time} are discretized in a way that preserves the locality of the Lagrangian. The time discretization (in units of $\tau$) pushes the second Dirac point up to energies of order $\hbar/\tau$, where it does not affect the low-energy physics --- as we demonstrate by comparing quantum Monte Carlo simulations with results from bosonization \cite{Hal81,Gia03,Giu08,vanDelft98}.

The lattice fermion approach which we will now describe refers specifically to the massless Dirac fermions that appear in topological insulators. Other approaches exist that exploit the boson-fermion correspondence. One can first bosonize the fermion formulation of the problem \cite{Alv86} and then put it on a lattice \cite{Ber23}. Luttinger liquid physics may also govern the low-energy properties of bosonic systems such as spin chains \cite{Hal81b}, where fermion doubling does not apply and a lattice formulation poses no difficulties \cite{Alc92,Uri21}.

{\em Locally discretized Lagrangian ---}
We construct the space-time lattice using the tangent fermion discretization approach \cite{Sta82,Ben83,Two08,Pac21,Don22,Bee23}. We first outline that approach for the noninteracting case, in a Lagrangian formulation that is a suitable starting point for the interacting problem.

Consider a 1D free massless fermion field $\psi_\sigma(x,t)$ with Lagrangian density given by
\begin{equation}
{\cal L}_{\text{continuum}}=\textstyle{\sum_{\sigma}}\psi^\dagger_\sigma(i\partial_t+i\sigma v_{\rm F}\partial_x)\psi_\sigma.\label{Lcontinuum}
\end{equation}
The spin degree of freedom $\sigma$, equal to $\uparrow\downarrow$ or $\pm 1$, distinguishes right-movers from left-movers, both propagating with velocity $v_{\rm F}$ along the $x$-axis. We set $\hbar=1$ and denote partial derivatives by $\partial_x,\partial_t$. The chemical potential is set to to zero (the Dirac point, corresponding to a half-filled band).

We discretize space $x$ and time $t$, in units of $a$ and $\tau$, respectively. The naive discretization of space replaces $\partial_x\mapsto (2a)^{-1}(e^{a\partial_x}-e^{-a\partial_x})$, which amounts to $\partial_x f(x)\mapsto (2a)^{-1}[f(x+a)-f(x-a)]$. Similarly, $\partial_t\mapsto (2\tau)^{-1}(e^{\tau\partial_t}-e^{-\tau\partial_t})$, producing a Lagrangian with a sine kernel,
\begin{equation}
{\cal L}_{\text{sine}}=(a\tau)^{-1}\textstyle{\sum_{\sigma}}\psi^\dagger_\sigma(\sin\hat{\omega}\tau-\sigma \gamma\sin\hat{k}a)\psi_\sigma.\label{Lsinedef}
\end{equation}
We defined the frequency and momentum operators $\hat{\omega}=i\partial_t$ and $\hat{k}=-i\partial_x$ and denote $\gamma=v_{\rm F}\tau/a$. The discretized $\psi$'s are dimensionless.

\begin{figure}[tb]
\centerline{\includegraphics[width=0.8\linewidth]{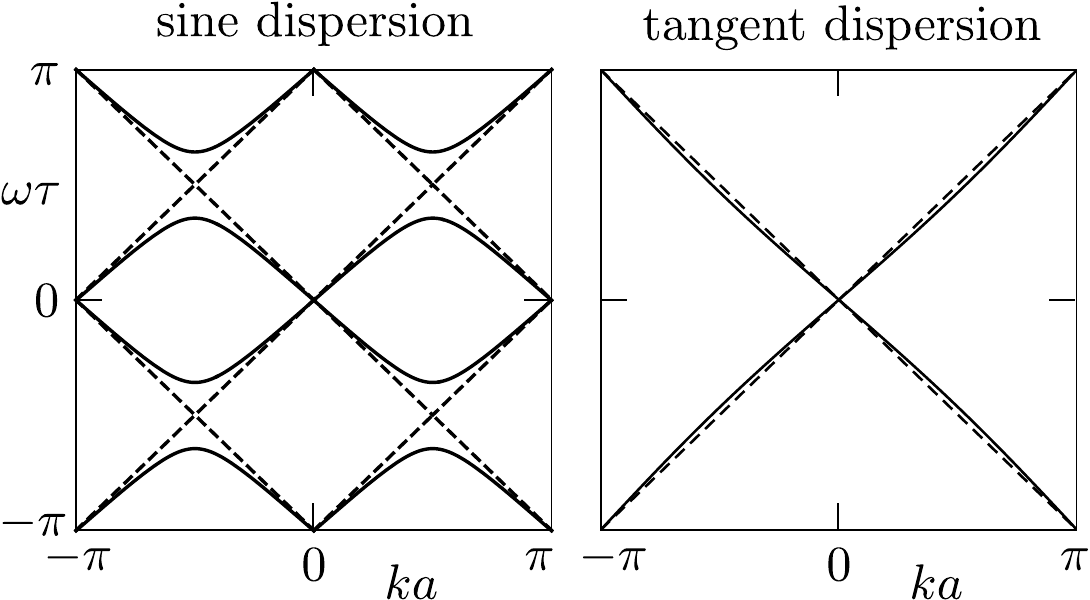}}
\caption{Dispersion relation of a massless fermion on a 1+1-dimensional space-time lattice. The two panels compare the sine and tangent discretization schemes, for $\gamma=v_{\rm F}\tau/a$ equal to 1 (dashed curves) or 0.9 (solid curves). The sawtooth discretization has the $\gamma$-independent dispersion $\omega=\pm v_{\rm F}\tau$ in the Brillouin zone $|\omega \tau|,|ka|<\pi$. Only the tangent discretization gives a local Lagrangian with a single Dirac point at $\omega=0$.
}
\label{fig_dispersion}
\end{figure}

The naive discretization is a local discretization, in the sense that the Lagrangian only couples nearby sites on the space-time lattice. However, it suffers from fermion doubling: The dispersion relation $\sin{\omega}\tau=\sigma \gamma\sin{k}a$ has branches of right-movers and left-movers which intersect at a Dirac point (see Fig.\ \ref{fig_dispersion}, left panel). Kramers degeneracy protects the crossings at time-reversally invariant points $\omega\tau,ka=0$ modulo $\pi$. In the Brillouin zone $|ka|,|\omega\tau|<\pi$ there are 4 inequivalent Dirac points, two of which are at $\omega=0$: one at $k=0$, the other at $|k|=\pi$. Low-energy scattering processes can couple these two Dirac points and open a gap without violating Kramers degeneracy. To avoid this we need to ensure that there is only a single Dirac point at $\omega=0$. 

One way to remove the spurious second species of low-energy excitations goes by the name of {\sc slac} fermions in the particle physics context \cite{Dre76}, or Floquet fermions in the context of periodically driven atomic lattices \cite{Bud17,Sun18}. In that approach one truncates the continuum linear dispersion at the Brillouin zone boundaries, and then repeats sawtooth-wise \cite{note1} with $2\pi$-periodicity:
\begin{equation}
{\cal L}_{\text{sawtooth}}=-i(a\tau)^{-1}\textstyle{\sum_{\sigma}}\psi^\dagger_\sigma(\ln e^{i\hat{\omega}\tau}-\sigma\gamma\ln e^{i\hat{k}a})\psi_\sigma.
\end{equation}
The sawtooth dispersion relation $\ln e^{i{\omega}\tau}=\sigma\gamma\ln e^{i{k}a}$ is strictly linear in the Brillouin zone, with a single Dirac point at $\omega=0$, however the Lagrangian is nonlocal:
\begin{equation}
(\ln e^{i\hat{k}a})f(x)= \sum_{n=1}^\infty (-1)^n n^{-1}[f(x-na)-f(x+na)],
\end{equation}
so distant points on the space-time lattice are coupled.

To obtain a local Lagrangian with a single Dirac point at $\omega=0$ we take two steps: First we replace the sine in ${\cal L}_{\text{sine}}$ by a tangent with the same $2\pi$ periodicity:
\begin{equation}
{\cal L}_{\text{tangent}}=\frac{2}{a\tau}{\textstyle\sum_{\sigma}}\psi^\dagger_\sigma\bigl[\tan(\hat{\omega}\tau/2)-\sigma \gamma\tan(\hat{k}a/2)\bigr]\psi_\sigma.\label{Ltangentpsi}
\end{equation}
The resulting tangent dispersion $\tan(\omega\tau/2)=\sigma\gamma\tan(ka/2)$ removes the spurious Dirac point (see Fig.\ \ref{fig_dispersion}, right panel), but it creates a non-local coupling. The locality is restored by the substitution
\begin{equation}
\psi_\sigma=\hat{D}\phi_\sigma,\;\;\hat{D}=\tfrac{1}{4}(1+e^{i\hat{k}a})(1+e^{i\hat{\omega}\tau}),\label{Phidef}
\end{equation}
which produces the Lagrangian
\begin{align}
{\cal L}_{\text{tangent}}={}&\tfrac{1}{2}(a\tau)^{-1}\textstyle{\sum_{\sigma}}\phi^\dagger_\sigma\bigl[(1+\cos\hat{k}a)\sin\hat{\omega}\tau\nonumber\\
&-\sigma \gamma(1+\cos\hat{\omega}\tau)\sin\hat{k}a\bigr]\phi_\sigma.\label{Ltangentphi}
\end{align}
Product terms $\cos\hat{k}a\times\sin\hat{\omega}\tau$ and $\cos\hat{\omega}\tau\times\sin\hat{k}a$ couple  $\phi_\sigma(x,t$) to $\phi_\sigma(x\pm a,t\pm\tau)$, so the coupling is off-diagonal on the space-time lattice but local.

This recovery of a local Lagrangian from a nonlocal Hamiltonian can be understood intuitively \cite{Bee23}: While the tangent discretization of the differential operator is nonlocal, its functional inverse, which is the trapezoidal integration rule, is local --- allowing for a local path integral formulation of the quantum dynamics.

The next step is to introduce the on-site Hubbard interaction (strength $U$, repulsive for $U>0$, attractive for $U<0$) by adding to ${\cal L}_{\text{tangent}}$ the term
\begin{equation}
{\cal L}_{\text{Hubbard}}=-(U/a)n_\uparrow(x,t)n_{\downarrow}(x,t),\;\;
n_\sigma=\,:\!\psi_\sigma^\dagger\psi_\sigma^{\vphantom{\dagger}}\!:\,
\label{LHubbard}
\end{equation}
The density $n_\sigma$ is normal ordered (Fermi sea expectation value is subtracted). 
Substitution of Eq.\ \eqref{Phidef} expresses the density $n_\sigma$ at point $(x,t)$ in terms of the average of the field $\phi_\sigma$ over the four corners of the adjacent space-time unit cell.

This completes the lattice formulation of the Luttinger liquid. We characterize it by the functions
\begin{align}
&C_\sigma(x)=\langle\psi_\sigma^\dagger(x,0)\psi_\sigma^{\vphantom{\dagger}}(0,0)\rangle,\;\;\bm{\psi}=(\psi_\uparrow,\psi_\downarrow),
\label{Csigmaxxdef}\\
&R_x(x)=\langle \rho_x(x)\rho_x(0)\rangle,\;\;\rho_x(x)=\tfrac{1}{2}\bm{\psi}^\dagger(x,0)
\bm{\sigma}_x\bm{\psi}(x,0).\nonumber
\end{align}
Here $\langle\cdots\rangle=Z^{-1}\operatorname{Tr}e^{-\beta H}\cdots$ indicates the thermal average at inverse temperature $\beta=1/k_{\rm B}T$ (with $Z=\operatorname{Tr}e^{-\beta H}$ the partition function). We first focus on the propagator $C_\sigma$.

\textit{Discretized Euclidean action ---}
The propagator can be rewritten as a fermionic path integral \cite{Mahan,AltlandSimons} over anticommuting fields $\Psi=\{\Psi_+,\Psi_-\}$ and $\bar{\Psi}=\{\bar{\Psi}_+,\bar{\Psi}_-\}$,
\begin{equation}
C_\sigma(x)=Z^{-1}\int{\cal D}\bar{\Psi}\int{\cal D}\Psi\,e^{-{\cal S}[\Psi,\bar{\Psi}]}\bar{\Psi}_\sigma(x,0)\Psi_\sigma(0,0),\label{Csigmadef}
\end{equation}
with ${\cal S}$ the Euclidean action. For free fermions one has
\begin{equation}
{\cal S}=\int_0^\beta dt \int_{0}^L dx\,\sum_\sigma\bar{\Psi}_\sigma(x,t)(\partial_t-i\sigma v_{\rm F}\partial_x)\Psi_\sigma(x,t).
\end{equation}
The Lagrangian \eqref{Lcontinuum} is integrated along the interval $0<it<i\beta$ on the imaginary time axis, with antiperiodic boundary conditions: $\Psi_\sigma(x,\beta)=-\Psi_\sigma(x,0)$. On the real space axis the integral runs from $0$ to $L$ with periodic boundary conditions, $\Psi_\sigma(0,t)=\Psi_\sigma(L,t)$.

The tangent fermion discretization replaces $i\partial_t\mapsto (2/\tau)\tan(\hat{\omega}\tau/2)$ and $i\partial_x\mapsto -(2/a)\tan(\hat{k}a/2)$, resulting in the discretized Euclidean action
\begin{subequations}
\begin{align}
{\cal S}_{\rm tangent}={}&2\textstyle{\sum_{x,t,\sigma}}\bar{\Psi}_\sigma(x,t)\bigl(-i\tan(\hat{\omega}\tau/2) \nonumber\\
&+\gamma\sigma\tan(\hat{k}a/2)\bigr)\Psi_\sigma(x,t)
\label{StangentPsi}\\
={}&\tfrac{1}{2}\textstyle{\sum_{x,t,\sigma}}\bar{\Phi}_\sigma(x,t)\bigl(-i(1+\cos\hat{k}a)\sin\hat{\omega}\tau \nonumber\\
&+\gamma\sigma(1+\cos\hat{\omega}\tau)\sin\hat{k}a\bigr)\Phi_\sigma(x,t).
\end{align}
\label{Stangent}
\end{subequations}
In the second equality we substituted the locally coupled fields, $\Psi=\hat{D}\Phi$, $\bar{\Psi}=\bar{\Phi}\hat{D}^\dagger$, cf.\ Eq.\ \eqref{Phidef}. The Hubbard interaction is then included by adding to ${\cal S}_{\rm tangent}$ the action
\begin{equation}
{\cal S}_{\rm Hubbard}=U\tau\textstyle{\sum_{x,t}} \bar{\Psi}_+(x,t)\Psi_+(x,t)\bar{\Psi}_-(x,t)\Psi_-(x,t).\label{SHubbard}
\end{equation}

We choose discretization units $\tau,a$ so that both $\beta/\tau$ and $L/a$ are integer. The space-time lattice consists of the points $it_n=in\tau$, $n=0,1,2\ldots \beta/\tau-1$, on the imaginary time axis and $x_n=na$, $n=0,1,2\ldots L/a-1$ on the real space axis. Upon Fourier transformation the sum over $t_n$ becomes a sum over the Matsubara frequencies $\omega_n=(2n+1)\pi/\beta$, while the sum over $x_n$ becomes a sum over the momenta $k_n=2n\pi/L$. These are odd versus even multiples of the discretization unit, to ensure the antiperiodic versus periodic boundary conditions in $t$ and $x$, respectively. In order to avoid the pole in the tangent dispersion we choose $\beta/\tau$ even and $L/a$ odd.

\textit{Free-fermion propagator ---}
Without the interaction term the propagator \eqref{Csigmadef} is given by a Gaussian path integral \cite{Mahan,AltlandSimons}, which evaluates to
\begin{equation}
C_\sigma(x)=\frac{\tau}{\beta L}\sum_{k,\omega}\frac{e^{-ikx}}{2i\tan({\omega}\tau/2)-2\gamma\sigma\tan({k}a/2)}.
\end{equation}
A simple closed-form answer follows for the Fourier transform $C_\sigma(k)$ in the zero-temperature $(\beta\rightarrow\infty$) limit,
\begin{align}
C_\sigma(k)={}&\tau\int_{-\pi/\tau}^{\pi/\tau}\frac{d\omega}{2\pi}\,\frac{1}{2i\tan({\omega}\tau/2)-2\gamma\sigma\tan({k}a/2)}\nonumber\\
={}&\frac{-1}{2\operatorname{sign}(\sigma\tan(ka/2))+2\gamma\sigma\tan(ka/2)}.\label{Cktangent}
\end{align}

For the sine dispersion we have instead
\begin{align}
C_\sigma(k)
={}&\frac{-\operatorname{sign}(\sigma\sin ka)}{\sqrt{1+4\gamma^2\sin^2 ka}},\label{Cksine}
\end{align}
while the sawtooth dispersion gives
\begin{align}
C_\sigma(k)
={}&-\frac{1}{\pi}\arctan\left(\frac{\pi}{\gamma\sigma ka}\right),\;\;|ka|<\pi.\label{Cksawtooth}
\end{align}

\begin{figure}[tb]
\centerline{\includegraphics[width=0.8\linewidth]{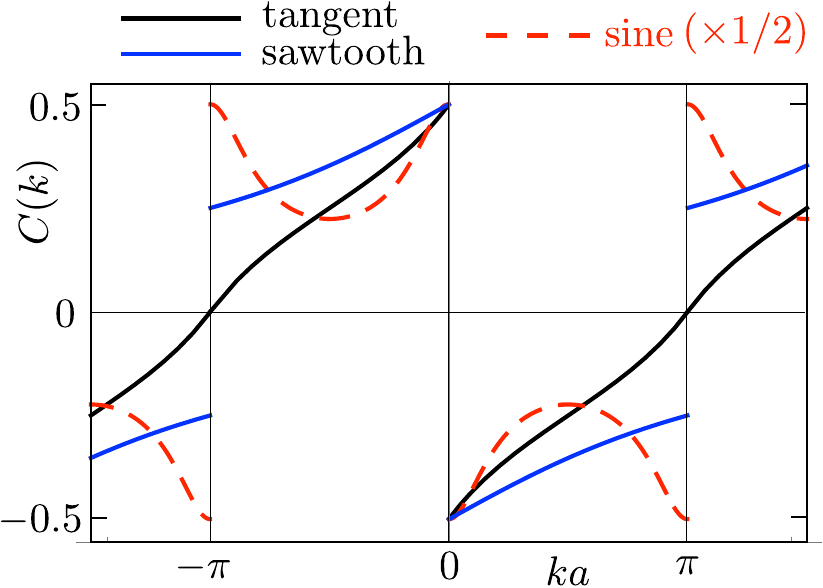}}
\caption{Free-fermion propagator in momentum space at zero temperature, calculated for three different discretization schemes. The plots follow from Eqs.\ \eqref{Cktangent}, \eqref{Cksine}, and \eqref{Cksawtooth}, for $\gamma=1$, $\sigma=+1$. Only the tangent fermion discretization is continuous at the Brillouin zone boundary $ka=\pm\pi$.
}
\label{fig_correlator}
\end{figure}

\begin{figure}[tb]
\centerline{\includegraphics[width=0.8\linewidth]{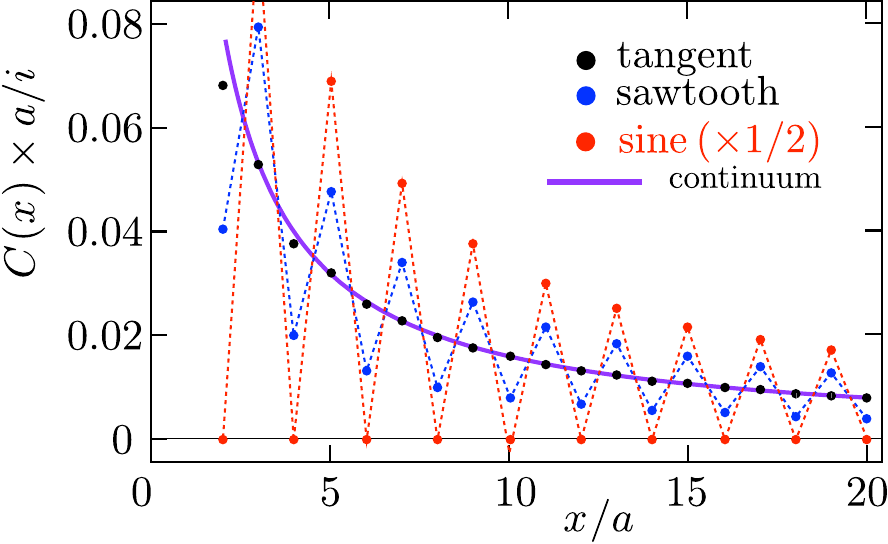}}
\caption{Same as Fig.\ \ref{fig_correlator}, but now in real space. The continuum result at zero temperature is $C(x)=i/2\pi x$ (solid curve), close to the tangent fermion discretization (black dots). The dashed lines are guides to the eye, to highlight the oscillatory behavior of the sawtooth and sine discretizations.
}
\label{fig_Cx}
\end{figure}

\begin{figure*}[tb]
\centerline{\includegraphics[width=0.8\linewidth]{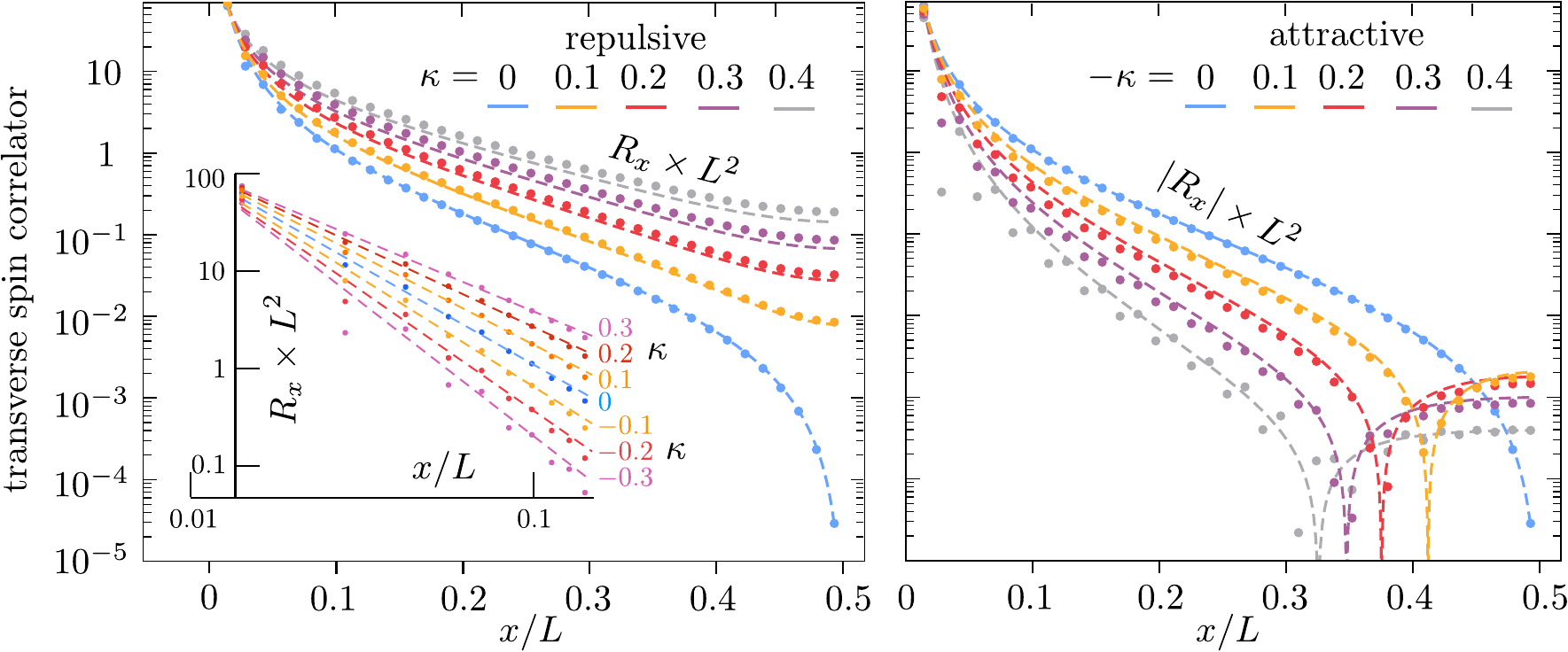}}
\caption{Main panels: The data points show the quantum Monte Carlo results for the correlator $R_x(x)=\tfrac{1}{4}\langle\bm{\psi}^\dagger(x)
\bm{\sigma}_x\bm{\psi}(x)\bm{\psi}^\dagger(0)
\bm{\sigma}_x\bm{\psi}(0)\rangle$ of the helical Luttinger liquid, on the space-time lattice with parameters $\beta/\tau=34$, $L/a=71$, $v_{\rm F}=a/\tau$. The different colors refer to different Hubbard interaction strengths $\kappa=Ua/2\pi v_{\rm F}$, repulsive on the left panel and attractive on the right panel. In the latter case the correlator $R_x$ changes sign, the plot shows the absolute value on a log-linear scale. The $x$-dependence at $x$ and $L-x$ is the same, because of the periodic boundary conditions, so only the range $0<x<L/2$ is plotted. The numerical data on the lattice is compared with the analytical bosonization theory in the continuum (dashed curves \cite{suppl}). Note that the lattice calculation slightly overestimates the interaction strength, for both the repulsive and attractive cases. The inset in the left panel combines data for both repulsive and attractive interactions on a log-log scale, to compare with the power law decay \eqref{powerlaw} (dashed lines).
}
\label{fig_MCresults}
\end{figure*}

Each dispersion has the expected continuum limit \cite{note2} $C_\sigma(k)\rightarrow-\tfrac{1}{2}\operatorname{sign}(\sigma k)$ for $|ka|\ll 1$, up to a factor of two for the sine dispersion due to fermion doubling. The difference appears near the boundary $|ka|= \pi/a$ of the Brillouin zone. As shown in Fig.\ \ref{fig_correlator}, only the tangent dispersion gives a propagator that is continuous across the Brillouin zone boundary. In real space, the discontinuity shows up as an oscillation of $C_\sigma(x)$ for separations $x$ that are even or odd multiples of $a$, see Fig.\ \ref{fig_Cx}. This is a known artefact of a finite band width \cite{Fer95} which is avoided by tangent fermions: their $C_\sigma(x)$ is close to the continuum result $i/2\pi x$ for $x$ larger than a few lattice spacings.

It is essential that the spatial discretization is accompanied by a discretization of (imaginary) time: If we would only discretize space, taking the limit $\tau\rightarrow 0$ at fixed $a$, then $\gamma\rightarrow 0$ and the propagator tends to the wrong limit,
\begin{equation}
\lim_{\tau\rightarrow 0}C_\sigma(x)=\tfrac{1}{2}i\sigma\int_0^{\pi/a}\sin kx \,dk=\frac{i\sigma\sin^2(\pi x/2a)}{\pi x},
\end{equation}
irrespective of how space is discretized. This deficiency of the sawtooth ({\sc slac}) approach was noted in Ref.\ \onlinecite{Wan23}.

\textit{Luttinger liquid correlators --}
We now include the Hubbard interaction \eqref{SHubbard} in the discretized Euclidean action \eqref{Stangent}, and evaluate the path integral \eqref{Csigmadef}  numerically by the quantum Monte Carlo method \cite{suppl}. In a Luttinger liquid the zero-temperature correlators decay as a power law \cite{Gia03},
\begin{subequations}
\begin{align}
&C_\sigma^2\propto x^{-K-1/K},\;\;R_x\propto x^{-2K},\\
&K=\sqrt{(1-\kappa)/(1 +\kappa)},\;\;\kappa=\frac{Ua}{2\pi v_{\rm F}}\in(-1,1).
\end{align}\label{powerlaw}
\end{subequations}
For repulsive interactions, $U>0\Rightarrow K<1$, the transverse spin-density correlator $R_x$ decays more slowly than the $1/x^2$ decay expected from a Fermi liquid.

Results for the interaction dependent decay are shown in Fig.\ \ref{fig_MCresults}. The data from the quantum Monte Carlo calculation of $R_x(x)$ is compared with the predictions from bosonization theory \cite{vanDelft98}. The power law decay \eqref{powerlaw} applies to an infinite 1D system. For a more reliable comparison with the numerics we include finite size effects in the bosonization calculations \cite{suppl}. 

The finite band width $1/\tau$ on the lattice requires that the dimensionless interaction strength $\kappa$ is small compared to unity. As we see in Fig.\ \ref{fig_MCresults} the agreement with the continuum results (dashed curves) remains quite satisfactory for $|\kappa|$ up to about $0.4$. We stress that this comparison does not involve any adjustable parameter.

\textit{Conclusion ---} We have shown that it is possible to faithfully represent an interacting Luttinger liquid on a lattice, without compromising the fundamental symmetries of massless fermions. The key step is a space-time discretization of the Lagrangian which is \textit{local} but does not introduce a spurious second species of low-energy excitations. We have tested the validity of this ``tangent fermion'' approach in the simplest setting where we can compare with the known bosonization results in the continuum. 

We anticipate that tangent fermions can become a powerful tool for the study of topological states of matter, where it is essential to maintain the topological protection of an unpaired Dirac cone. An application to the fermionic Casimir effect was published in Ref.\ \onlinecite{Bee24}. We have shown that the technique can be applied to quantum Monte Carlo calculations, but we expect it to be more generally applicable to fermionic lattices. Indeed, a second quantized formulation has very recently been used to avoid fermion doubling in the context of tensor networks \cite{Hae24}.

\textit{Acknowledgments ---} C.B. received funding from the European Research Council (Advanced Grant 832256). J.T. received funding from the National Science Centre, Poland, within the QuantERA II Programme that has received funding from the European Union's Horizon 2020 research and innovation programme under Grant Agreement Number 101017733, Project Registration Number 2021/03/Y/ST3/00191, acronym {\sc tobits}.



\appendix

\makeatletter

\section{Derivation of the Euclidean action in the tangent discretization}

In this appendix, we explain how to arrive at the Euclidean action given in Eqs.\ \eqref{Stangent} and \eqref{SHubbard}. In particular, we show that the shift in chemical potential, originating from the normal ordering of the density operator in the Hubbard-interaction Lagrangian \eqref{LHubbard}, is absent in the Euclidean action. We start from the continuum Hamiltonian $H = H_0 + H_\text{Hubbard}$, with
\begin{align}
H_0 ={}& \sum_{\sigma,k} \sigma v_\text{F} k :\!\psi^\dag_\sigma(k) \psi_\sigma(k)\!: \nonumber\\
={}& \sum_{\sigma,k} \sigma v_\text{F} k 
\left[\psi^\dag_\sigma(k)\psi_\sigma(k) - 1 + \theta(\sigma k)\right],\\
H_\text{Hubbard} ={}& U \int d x\, n_+(x) n_-(x), \\
n_\sigma(x) ={}& :\!\psi_\sigma(x)^\dag \psi_\sigma(x)\!:\, \\
=\frac{1}{L}\sum_{k\neq k'} &e^{i (k-k')x} [\psi_\sigma(k')^\dag \psi_\sigma(k) 
 + (\theta(\sigma k)-1)\delta_{kk'}],\nonumber
\end{align}
where the normal ordering $:\dots:$ brings to the right operators $\psi_\sigma(k)$ with $\sigma k>0$, and $\psi^\dag_\sigma(k)$ with $\sigma k\leq 0$. Here $\theta$ is the unit step function, defined by $\theta(\sigma k)=1$ for $\sigma k>0$, and $\theta(\sigma k)=0$ for $\sigma k\leq 0$. 

We introduce fermionic coherent states $|\Psi\rangle$, which satisfy
\begin{equation}
\psi_\sigma(k)|\Psi\rangle = \Psi_\sigma(k)|\Psi\rangle, \;\;
\langle\Psi|\psi_\sigma^\dag(k) = \langle\Psi|\bar\Psi_\sigma(k) \,.
\end{equation}
Then, with $\tau=\beta/N$ and $|\Psi(0)\rangle = -|\Psi(N)\rangle$, where $N$ is a positive integer, the partition function is
\begin{align}
Z = \operatorname{Tr} e^{-\beta H} ={}& \lim_{N\rightarrow\infty}
\int \prod_{t=1}^{N} \mathcal{D} \Psi(t) \mathcal{D} \bar\Psi(t) \nonumber\\
&
e^{-\bar\Psi(t)\Psi(t)}
\langle\Psi(t)|e^{-\tau H}|\Psi(t-1)\rangle \,,
\end{align}
where $\Psi(0) = -\Psi(N)$. 

We perform a Hubbard-Stratonovich transformation,
\begin{widetext}
\begin{equation}
e^{-\tau H_\text{Hubbard}}
=
\int \mathcal{D}W(t) \mathcal{D}W^*(t) 
\exp\Bigg\{
\int d x -\frac{\tau|W(x,t)|^2}{|U|}
+ \tau W(x,t) n_+(x) - \tau \operatorname{sign}(U) W^*(x,t) n_-(x)
\Bigg\}\,,
\end{equation}
which yields
\begin{multline}
Z =
\lim_{N\rightarrow\infty}
\int \left(\prod_{t=1}^{N} \mathcal{D} \Psi(t) \mathcal{D} \bar\Psi(t) 
\mathcal{D}W(t) \mathcal{D}W^*(t) \right) 
\exp\Bigg\{\sum_t\Bigg[
-\int d x\, \frac{\tau|W(x,t)|^2}{|U|}
-\bar\Psi(t)[\Psi(t)-\Psi(t-1)]
\\
-\tau\sum_{\sigma,k} \sigma v_\text{F} k
[\bar\Psi_\sigma(k,t) \Psi_\sigma(k,t-1)
-1 + \theta(\sigma k)
]
\\
+ \frac{\tau}{L}\int d x\, W(x,t) 
\sum_{k,k'} e^{i (k-k')x}
[\bar\Psi_+(k',t) \Psi_+(k,t-1) + ( \theta(k)-1)\delta_{kk'}] \\
- \frac{\tau}{L}\operatorname{sign}(U) \int d x\, W^*(x,t) 
\sum_{kk'} e^{i (k-k')x}
[\bar\Psi_-(k',t) \Psi_-(k,t-1) + (-1 + \theta(-k))\delta_{kk'}] 
\Bigg]\Bigg\} \,.
\end{multline}

We make the variable change
\begin{equation}
\Psi_\sigma(k,t) \mapsto \frac{1}{2}[\Psi_\sigma(k,t) + \Psi_\sigma(k,t+1)] \,,
\end{equation}
under which the partition function becomes
\begin{multline}
Z =
\lim_{N\rightarrow\infty}
\int \left(\prod_{t=1}^{N} \mathcal{D} \Psi(t) \mathcal{D} \bar\Psi(t) 
\mathcal{D}W(t) \mathcal{D}W^*(t) \right) 
\exp\Bigg\{\sum_t\Bigg[
-\int d x \frac{\tau|W(x,t)|^2}{|U|} \\
-\frac{1}{2}\sum_\sigma \sum_{k,k'} \left(\delta_{kk'} + \tau\Lambda_\sigma(k,k')\right)
\bar\Psi_\sigma(k',t)[\Psi_\sigma(k,t+1) - \Psi_\sigma(k,t-1)] 
\\
-\tau\sum_\sigma
\sum_{k} \sigma v_\text{F} k
\left(
 \bar\Psi_\sigma(k,t) \frac{\Psi_\sigma(k,t-1) + 2\Psi_\sigma(k,t) + \Psi_\sigma(k,t+1)}{4} 
-1 + \theta(\sigma k)
\right)
\\
+ \frac{\tau}{L}\int d x\, W(x,t) 
\sum_{k,k'} e^{i (k-k')x}
\left(\bar\Psi_+(k',t) \frac{\Psi_+(k,t+1) + 2\Psi_+(k,t) + \Psi_+(k,t-1)}{4}
+ ( \theta(k)-1)\delta_{kk'}
\right)
 \\
- \frac{\tau}{L}\operatorname{sign}(U) \int d x\, W^*(x,t) 
\sum_{k,k'} e^{i (k-k')x}
\left(\bar\Psi_-(k',t) \frac{\Psi_-(k,t+1) + 2\Psi_-(k,t) + \Psi_-(k,t-1)}{4}
+ ( \theta(-k)-1)\delta_{kk'}
\right)
\Bigg]\Bigg\} \,,
\end{multline}
where
\begin{equation}
\Lambda_\sigma(k,k') = -\frac{\sigma v_\text{F} k}{2}\delta_{kk'}
+ \frac{1}{2 L}\int d x \left[W(x,t)e^{i (k-k')x}\delta_{\sigma+} - \operatorname{sign}(U) W^*(x,t)e^{i (k-k')x}\delta_{\sigma-}\right]\,.
\end{equation}

We next make the substitution
\begin{equation}
\bar \Psi_\sigma(k') \mapsto \sum_{k''} [\delta_{k'k''} - \tau\Lambda_\sigma(k',k'')]\Psi_\sigma(k'')\,.
\end{equation}
Ignoring the subleading terms, the Jacobian of the variable change is
\begin{equation}
\det(\delta_{k'k''} - \tau\Lambda_\sigma(k',k''))
=
\exp\left\{-\tau \sum_{k} 
\left(
-\sigma v_\text{F} k
+ \frac{1}{2 L}\int d x \left[W(x,t) - \operatorname{sign}(U) W^*(x,t)\right]
\right)
\right\}\,.
\end{equation}
The change of measure of Grassmann variables produces the inverse of the Jacobian, which cancels the shift in chemical potential [the term $ \theta(\sigma k)-1$] from normal ordering. This yields the action
\begin{multline}
Z =
\lim_{N\rightarrow\infty}
\int \left(\prod_{t=1}^{N} \mathcal{D} \Psi(t) \mathcal{D} \bar\Psi(t) 
\mathcal{D}W(t) \mathcal{D}W^*(t) \right) 
\exp\Bigg\{\sum_t\Bigg[
-\int d x \frac{\tau|W(x,t)|^2}{|U|} \\
-\frac{1}{2}\sum_{\sigma,k}\bar\Psi_\sigma(k,t)[\Psi_\sigma(k,t+1) - \Psi_\sigma(k,t-1)] 
\\
-\tau\sum_{\sigma,k} \sigma v_\text{F} k
 \bar\Psi_\sigma(k,t) \frac{\Psi_\sigma(k,t-1) + 2\Psi_\sigma(k,t) + \Psi_\sigma(k,t+1)}{4} 
\\
+ \tau\int d x\, W(x,t) 
\bar\Psi_+(x,t) \frac{\Psi_+(x,t+1) + 2\Psi_+(x,t) + \Psi_+(x,t-1)}{4}
 \\
- \tau\operatorname{sign}(U) \int d x W^*(x,t) 
\bar\Psi_-(x,t) \frac{\Psi_-(x,t+1) + 2\Psi_-(x,t) + \Psi_-(x,t-1)}{4}
\Bigg]\Bigg\} \,.
\end{multline}

Finally we can integrate out the Hubbard-Stratonovich field, substitute $\Psi_\sigma \mapsto 2(1 + \cos\hat\omega\tau)^{-1}\Psi_\sigma$, and apply the tangent discretization $k\mapsto 2a^{-1}\tan(ak/2)$, which results in
\begin{multline}
Z =
\lim_{N\rightarrow\infty}
\int \left(\prod_{t=1}^{N} \mathcal{D} \Psi(t) \mathcal{D} \bar\Psi(t) 
\right) 
\exp\Bigg\{-\sum_t\Bigg[
2\sum_{\sigma, k}
\bar\Psi_\sigma(k) (-i\tan(\hat\omega\tau/2)\Psi_\sigma(k) 
+\tau \sigma v_\text{F} \tan(k/2)) \Psi_\sigma(k,t) \\
+ \tau U \sum_x \bar\Psi_+(x,t) \Psi_+(x,t) \bar\Psi_-(x,t) \Psi_-(x,t)
\Bigg]\Bigg\} \,.
\end{multline}
For finite $N$, this is the partition function corresponding to the tangent-discretized action given in Eqs.\ \eqref{Stangent} and \eqref{SHubbard} in the main text.

\end{widetext}

\section{Quantum Monte Carlo calculation}
\label{sec_QMC}

\subsection{Hubbard-Stratonovich transformation of the Euclidean action}

To evaluate the fermionic path integral representation of the partition function,
\begin{equation}
Z=\int{\cal D}\bar{\Psi}\int{\cal D}\Psi\,e^{-{\cal S}[\Psi,\bar{\Psi}]}\label{Zdef}
\end{equation}
we follow the usual auxiliary field approach \cite{Sca81,Hir83,Hir86,Gub16,Bec17}, by which the two-body Hubbard interaction is transformed into a sum of one-body terms coupled to a fluctuating Ising field $s(x,t)\in\{+1,-1\}$. In a Hamiltonian formulation this is accomplished by the discrete Hubbard-Stratonovich transformation of Ref.\ \onlinecite{Hir83}. We cannot follow that route, because the tangent fermion Hamiltonian is nonlocal, instead we need to work with the Lagrangian formulation --- which is local. 

Starting from the discretized Euclidean action in Eqs.\ \eqref{Stangent} and \eqref{SHubbard}
we factor out the two-body term,
\begin{equation}
e^{-{\cal S}}=e^{-{\cal S}_{\rm tangent}}\prod_{x,t}e^{-U\tau \bar{\Psi}_+(x,t)\Psi_+(x,t)\bar{\Psi}_-(x,t)\Psi_-(x,t)}.
\end{equation}
This is allowed because all bilinears $\bar{\Psi}\Psi$ of the anticommuting Grassmann fields commute. (The approximate Trotter splitting \cite{Gub16,Bec17} from the Hamiltonian formulation does not appear here.)

Focusing on one factor, we have the sequence of identities (using $\Psi^2=\bar{\Psi}^2=0$)
\begin{widetext}
\begin{align}
e^{-U\tau \bar{\Psi}_+\Psi_+\bar{\Psi}_-\Psi_-}={}&1-U\tau \bar{\Psi}_+\Psi_+\bar{\Psi}_-\Psi_-\nonumber\\
={}&\tfrac{1}{2}\sum_{s=\pm}\bigl[1+s\sqrt{|U\tau}|(\bar{\Psi}_+\Psi_+-\operatorname{sign}(U\tau)\bar{\Psi}_-\Psi_-)-U\tau s^2\bar{\Psi}_+\Psi_+\bar{\Psi}_-\Psi_-\bigr]\nonumber\\
={}&\tfrac{1}{2}\sum_{s=\pm}\exp\bigl[s\sqrt{|U\tau}|(\bar{\Psi}_+\Psi_+-\operatorname{sign}(U\tau)\bar{\Psi}_-\Psi_-)].\end{align}
Collecting all factors we thus arrive at the desired Hubbard-Stratonovich transformation of the Euclidean action,
\begin{equation}
e^{-{\cal S}}=e^{-{\cal S}_{\rm tangent}}\frac{1}{2^{\frac{\beta L}{\tau a}}}\sum_{s(x,t)=\pm 1}\exp\biggl[|U\tau|^{1/2}\sum_{x,t}s(x,t) \biggl(\bar{\Psi}_+(x,t)\Psi_+(x,t)-\operatorname{sign}(U\tau)\bar{\Psi}_-(x,t)\Psi_-(x,t)\biggr)\biggr],\label{HStr}
\end{equation}

In the tangent fermion discretization the charge density $\bar{\Psi}{\Psi}$ is rewritten in terms of the locally coupled fields $\Phi,\bar{\Phi}$, cf.\ Eq.\ \eqref{Phidef},
\begin{align}
\bar{\Psi}_\pm(x,t){\Psi}_\pm(x,t)={}&\bar{\Phi}_\pm\hat{D}^\dagger|x,t\rangle\langle x,t| \hat{D}{\Phi}_\pm=\tfrac{1}{16}\bar{\Phi}_\pm(1+e^{-i\hat{k}a})(1+e^{-i\hat{\omega}t})|x,t\rangle\langle x,t| (1+e^{i\hat{k}a})(1+e^{i\hat{\omega}t}){\Phi}_\pm,\nonumber\\
={}&\tfrac{1}{16}[\bar{\Phi}_\pm(x,t)+\bar{\Phi}_\pm(x-a,t)+\bar{\Phi}_\pm(x,t+\tau)+\bar{\Phi}_\pm(x-a,t+\tau)\bigr]\nonumber\\
&\times[\Phi_\pm(x,t)+\Phi_\pm(x-a,t)+\Phi_\pm(x,t+\tau)+\Phi_\pm(x-a,t+\tau)\bigr].
\end{align}
The Jacobian $J=\det \hat{D}^\dagger \hat{D}$ of the transformation is independent of the Ising field.

For any given Ising field configuration $s(x,t)$ the action is now quadratic in the Grassmann fields $\Phi,\bar{\Phi}$,
\begin{equation}
{\cal S}[\Phi,\bar{\Phi}, s]= \sum\limits_{\sigma,x,x',t,t'} \bar{\Phi}_\sigma(x',t')M_{\sigma}(x,x',t,t')[s]\Phi_\sigma(x,t),
\end{equation}
with a \textit{local} kernel
\begin{align}
M_\sigma(x,x',t,t')[s]={}&\tfrac{1}{2}\bigl(-i(1+\cos\hat{k}a)\sin\hat{\omega}\tau+\sigma\gamma(1+\cos\hat{\omega}\tau)\sin\hat{k}a\bigr)\nonumber\\
&+\delta_{x,x'}\delta_{t,t'}|U\tau|^{1/2}s(x,t)D^\dagger |x,t\rangle\langle x,t|D\times\begin{cases}
\sigma&\text{if}\;\;U>0,\\
1&\text{if}\;\;U<0.
\end{cases}\label{Msigmadef}
\end{align}
\end{widetext}

The Gaussian path integral over the fields $\Phi,\bar{\Phi}$ produces a weight functional
\begin{equation}
P[s]=\int{\cal D}\bar{\Phi}\int{\cal D}\Phi\,e^{-{\cal S}[\Phi,\bar{\Phi},s]}=\det M_{+}[s]M_{-}[s]
\end{equation}
for the average over the Ising field. This final average is carried out by means of the Monte Carlo importance sampling algorithm.

\subsection{Absence of a sign problem}

For the Monte Carlo averaging we need to ascertain the absence of a sign problem: The weight functional $P[s]$ should be non-negative for any Ising field configuration. This is indeed the case: From Eq.\ \eqref{Msigmadef} one sees that for the attractive interaction ($U<0$)
\begin{equation}
M^\ast_{-}[s] = M_{+}[s]\Rightarrow P[s] = |\det M_{+}[s] |^2. \label{sym_attr}
\end{equation}
(Note that $\hat{k}= -i\partial_x$ and $\hat\omega=i\partial_t$ changes sign upon complex conjugation.)  For the repulsive interaction ($U>0$)
\begin{align}
M^{\dagger}_{-}[s] = -M_{+}[s]\Rightarrow P[s] &=(-1)^{\beta L/\tau a} |\det M_{+}[s]|^2\nonumber\\
&=|\det M_{+}[s]|^2, \label{sym_rep}
\end{align}
because $\beta L/\tau a $ is an even integer.

\subsection{Correlators}

We apply the quantum Monte Carlo calculation to equal-time correlators of the form
\begin{widetext}
\begin{align}
&\langle \psi_{\sigma}^\dagger(x_1)\psi_{\sigma}(x_2)\psi_{\sigma'}^\dagger(x_3)\psi_{\sigma'}(x_4) \rangle=\langle (\hat{D}\phi_{\sigma})^\dagger(x_1)(\hat{D}\phi_{\sigma})(x_2)(\hat{D}\phi_{\sigma'})^\dagger(x_3)(\hat{D}\phi)_{\sigma'}(x_4) \rangle\nonumber\\
&\qquad=\int{\cal D}\bar{\Phi}\int{\cal D}\Phi\,\langle e^{-S[\Phi,\bar{\Phi},s]}\rangle_s (\bar{\Phi}_{\sigma}\hat{D}^\dagger)(x_1)(\hat{D}\Phi_{\sigma})(x_2)(\bar{\Phi}_{\sigma'}\hat{D}^\dagger)(x_3)(\hat{D}\Phi)_{\sigma'}(x_4)\nonumber\\
&\qquad=
\frac{1}{Z}\sum_{s(x,t)=\pm 1}P[s]\bigg( \hat{A}_{\sigma}(x_2,x_1)[s]\hat{A}_{\sigma'}(x_4,x_3)[s] - \delta_{\sigma\sigma'}\hat{A}_{\sigma}(x_4,x_1)[s]\hat{A}_{\sigma}(x_2,x_3)[s]\bigg). \label{corr_A}
\end{align}
In the final equality we defined $\hat{A}_{\sigma}[s] = \hat{D}M_{\sigma}^{-1}[s]\hat{D}^\dagger$ and we have used the integration formula (Wick's theorem)
\begin{equation}
\int{\cal D}\bar{\Phi}\int{\cal D}\Phi\,e^{-\bar{\Phi}M\Phi}\Phi_k\Phi_l\bar{\Phi}_m\bar{\Phi}_n=(\det M)\bigl[(M^{-1})_{kn}(M^{-1})_{lm}-(M^{-1})_{km}(M^{-1})_{ln}\bigr].
\end{equation}
\end{widetext}

We consider the spin correlator
\begin{align}
	R_x(x)={}&\tfrac{1}{4}\langle\bm{\psi}^\dagger(x)
	\bm{\sigma}_x\bm{\psi}(x)\bm{\psi}^\dagger(0)
	\bm{\sigma}_x\bm{\psi}(0)\rangle \nonumber\\
	={}&-\tfrac{1}{4}\big(\langle \psi_{\uparrow}^\dagger(x)
	\psi_\uparrow(0)\psi_{\downarrow}^\dagger(y)
	\psi_\downarrow(x)\rangle \nonumber\\
	& + \langle \psi_{\downarrow}^\dagger(x)
	\psi_\downarrow(0)\psi_{\uparrow}^\dagger(0)
	\psi_\uparrow(x)\rangle\big). \label{spincorr}
\end{align}
In the second equality we used spin conservation symmetry,
\begin{equation}
\begin{split}
\langle \psi_{\uparrow}^\dagger(x)
\psi_\downarrow(x)\psi_{\uparrow}^\dagger(0)
\psi_\downarrow(0)\rangle=0,\\
\langle \psi_{\downarrow}^\dagger(x)
\psi_\uparrow(x)\psi_{\downarrow}^\dagger(0)
\psi_\uparrow(0)\rangle=0.
\end{split}\label{spinconseq}
\end{equation}

We substitute Eq.\ \eqref{corr_A} into to Eq.\ \eqref{spincorr},
\begin{equation}
	R_x(x) = -\tfrac{1}{4}\big\langle \hat{A}_{\uparrow}(0,x)[s]\hat{A}_{\downarrow}(x,0)[s]+\hat{A}_{\downarrow}(0,x)[s]\hat{A}_{\uparrow}(x,0)[s]\bigr\rangle_s.
\end{equation}
Using the symmetries \eqref{sym_attr} and \eqref{sym_rep} of the $M$-matrix we simplify it to
\begin{equation}
	R_x(x) =\begin{cases}
		\tfrac{1}{4}\bigl\langle\big|A_{\uparrow}(x,0)[s]\big|^2+\big|A_{\uparrow}(0,x)[s]\big|^2\bigr\rangle_s&\text{if } U >0,\\
		-\tfrac{1}{2}\operatorname{Re}\bigl\langle\{A_{\uparrow}(x,0)[s]A^*_{\uparrow}(0,x)[s]\}\bigr\rangle_s&\text{if } U <0,
	\end{cases}\label{Rxfinal}
\end{equation}
with $\{\cdots\}$ the anticommutator.

The average in Eq.\ \eqref{Rxfinal} is over the Ising field $s$. To improve the statistics we make use of translational invariance in space and imaginary time, by additionally averaging the correlator over the initial position (replacing $(x,0)\mapsto(x+y,y)$ with $0<y<L$), as well as over $0<t<\beta$.

\subsection{Monte Carlo averaging}
For the Monte Carlo averaging we perform local updates of the auxiliary Ising field, one spin-flip at the time. An operator average is then sampled at each Monte Carlo iteration, which includes $\beta L/\tau a$ local spin-flip steps.

We have tried out three alternative numerical methods: 
\begin{enumerate}
\item The first method is a simple dense matrix calculation, where we recalculate $P[s]$ after each spin-flip and $A_{\uparrow}[s]$ after each Monte Carlo step, using neither the sparsity of the $M$-matrix nor the locality of the update. This is optimal for systems of small sizes $(L/a\lesssim 11,\;\beta/\tau\lesssim 6)$.
\item In the second method we use the locality of the update of the Ising field, by employing the Woodbury formula for the update of $P[s]$ and $A_{\uparrow}[s]$. 
This is favorable for systems of medium sizes $(L/a\approx 41,\,\beta/\tau\approx 20)$.
\item In the third method we use the sparsity of the $M$-matrix, with the help of the SuperLU library.\footnote{X. S. Li, \textit{An overview of SuperLU: Algorithms, implementation,              
and user interface, ACM Transactions on Mathematical Software}, \textbf{31}, 302 (2005).} This gives the best performance for large systems $(L/a \gtrsim 61,\,\beta/\tau \gtrsim 30)$.
\end{enumerate}

\begin{figure}[tb]
\centerline{\includegraphics[width=1\linewidth]{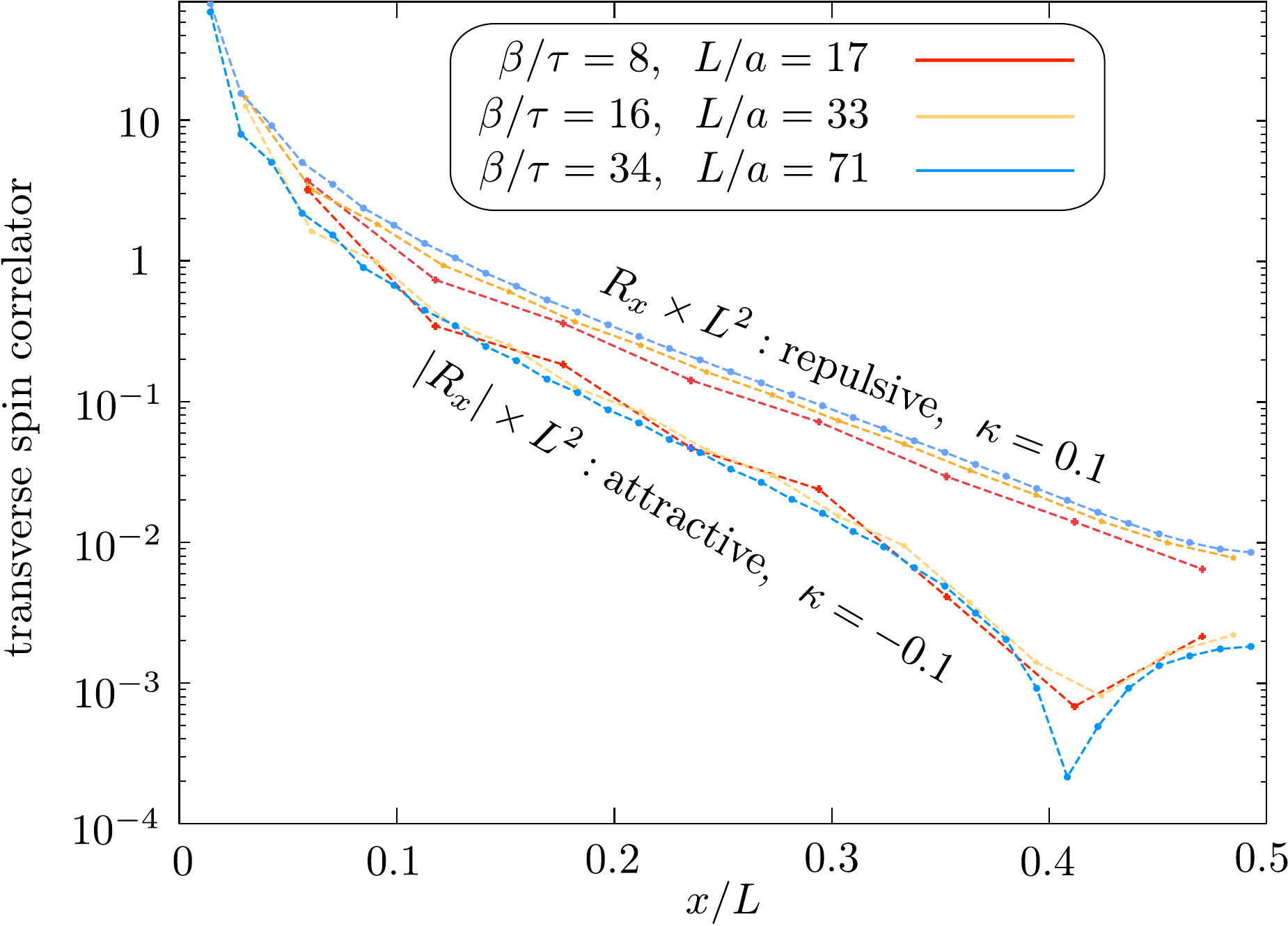}}
\caption{Quantum Monte Carlo results for the correlator $R_x(x)=\langle\bm{\psi}^\dagger(x)
\bm{\sigma}_x\bm{\psi}(x)\bm{\psi}^\dagger(0)
\bm{\sigma}_x\bm{\psi}(0)\rangle$ of the helical Luttinger liquid (interaction strength $|\kappa|=0.1$), on the space-time lattice for three different lattice sizes (at fixed $a/\tau=v_{\rm F}$). The data for the largest system corresponds to Fig.\ \ref{fig_MCresults} from the main text. The dotted line connecting the data points is a guide to the eye.
}
\label{fig_convergence}
\end{figure}

In Fig.\ \ref{fig_convergence} we compare results for three system sizes. We fix the ratio $a/\tau=v_{\rm F}$ of the discretization units of space and (imaginary) time and for each choice of $a,\tau$ we ensure that $L/a$ is an odd integer and $\beta/\tau$ is an even integer (to avoid the pole in the tangent dispersion). The size dependence is relatively weak for the repulsive interaction, and more significant for the attractive interaction near values of $x$ where the correlator changes sign.

\section{Bosonization results}
\label{sec_bosonization}

The curves in Fig.\ \ref{fig_MCresults} are the bosonization results for the spin correlator $R_x(x)$ of the helical Luttinger liquid on a ring of length $L$. We describe that calculation.

\subsection{Bosonic form of the Hamiltonian}

We start directly from the bosonic form of the Luttinger Hamiltonian \cite{vanDelft98},
\begin{align}
H ={} &\frac{2\pi v_{\rm F}}{L}\biggl[ \sum_{n_q=1}^\infty n_q \bigl(b^{\dagger}_{q\uparrow}b^{\vphantom{\dagger}}_{q\uparrow}+b^\dagger_{q\downarrow}b^{\vphantom{\dagger}}_{q\downarrow}\bigr)\nonumber\\
&-\kappa \sum_{n_q=1}^\infty n_q \bigl(b^{\dagger}_{q\uparrow}b^{\dagger}_{q\downarrow}+b_{q\uparrow}b_{q\downarrow}\bigr)\nonumber\\
&+\tfrac{1}{2} N_\uparrow(N_\uparrow+1) +\tfrac{1}{2}N_\downarrow(N_\downarrow+1)+\kappa N_\uparrow N_\downarrow \biggr].
\end{align}
The bosonic creation and annihilation operators $b^\dagger,b$ are constructed from the fermionic operators $c^\dagger,c$ by
\begin{align}
	b^\dagger_{q\sigma}=\frac{i}{\sqrt{n_q}}\sum_k c^\dagger_{ (k+q)\sigma}c_{ k\sigma}, \quad q = \frac{2\pi}{L}\sigma n_q.
\end{align}
We impose periodic boundary conditions, quantizing $k=2\pi n/L$, $n\in\mathbb{Z}$. The bosonic wave number $q=2\pi \sigma n_q/L$ is defined such that $n_q>0$ and the sign of $q$ is fixed by the spin index. 

The normally ordered fermionic number operator 
\begin{equation}
N_\sigma = \sum_k\colon\! c^\dagger_{k\sigma}c^{\vphantom{\dagger}}_{k\sigma}\!\colon=\sum_k \left(c^\dagger_{k\sigma}c^{\vphantom{\dagger}}_{k\sigma}-\langle 0|c^\dagger_{k\sigma}c^{\vphantom{\dagger}}_{k\sigma}|0\rangle\right)
\end{equation}
commutes with the bosonic operators. The Fermi sea expectation value $\langle 0|c^\dagger_{k\sigma}c^{\vphantom{\dagger}}_{k\sigma}|0\rangle$ is defined such that wave numbers $\sigma k\leq 0$ contribute, so all states with energy $E\leq 0$, including the spin-up and spin-down states at $k=0$. For later use we note that this implies that 
\begin{equation}
\langle N_\sigma\rangle=1/2-1=-1/2\label{Nonehalf}
\end{equation}
at zero temperature in a half-filled band (the state at $k=0$ only contributes $1/2$ per spin direction to the ground state).

The Hamiltonian can be diagonalised by a Bogoliubov transformation \cite{vanDelft98},
\begin{align}
	B_{q\pm} ={}&2^{-3/2}( K^{-1/2}+ K^{1/2})(b_{q\uparrow}\mp b_{q\downarrow}) \nonumber\\ 
	&\pm 2^{-3/2} ( K^{-1/2}- K^{1/2})(b^\dagger_{q\uparrow}\mp b^\dagger_{q\downarrow}),
\end{align}
with $K=\sqrt{1-\kappa}/\sqrt{1+\kappa}$, resulting in
\begin{align}
	H ={}&\frac{2\pi v_{\rm F}}{L}\biggl[\sqrt{1-\kappa^2} \sum_{n_q=1}^\infty n_q (B_{q+}^\dagger B_{q+}+B_{q-}^\dagger B_{q-})\nonumber\\
	& +\tfrac{1}{2}  N_\uparrow(  N_\uparrow + 1)+\tfrac{1}{2}  N_\downarrow(  N_\downarrow + 1)  + \kappa  N_{\uparrow} N_{\downarrow}\biggr].
\end{align}

\subsection{Spin correlator in terms of the bosonic fields}

We wish to compute the (equal-time) spin correlator
\begin{align}
	R_x(x)={}&\tfrac{1}{4}\langle\bm{\psi}^\dagger(x)
	\bm{\sigma}_x\bm{\psi}(x)\bm{\psi}^\dagger(0)
	\bm{\sigma}_x\bm{\psi}(0)\rangle \nonumber\\
	={}& -\tfrac{1}{2}\operatorname{Re}{\langle \psi_{\uparrow}^\dagger(x)
		\psi_\uparrow(0)\psi_{\downarrow}^\dagger(0)
		\psi_\downarrow(x)\rangle}.
\end{align}
In the second equality we used spin conservation symmetry \eqref{spinconseq}
and translational symmetry,
\begin{align}
	&\langle \psi_{\uparrow}^\dagger(x)
	\psi_\downarrow(x)\psi_{\downarrow}^\dagger(0)
	\psi_\uparrow(0)\rangle = \langle \psi_{\uparrow}^\dagger(0)
	\psi_\downarrow(0)\psi_{\downarrow}^\dagger(-x)
	\psi_\uparrow(-x)\rangle\nonumber\\
	& \qquad\qquad= \langle \psi_{\uparrow}^\dagger(0)
	\psi_\downarrow(0)\psi_{\downarrow}^\dagger(x)
	\psi_\uparrow(x)\rangle^*. \label{transsymeq}
\end{align}
The correlator diverges when $x\rightarrow 0$, we regularize this ultraviolet divergence with cutoff length $a_0$.

The fermion field is related to the bosonic operators by the ``refermionization'' relation \cite{vanDelft98}
\begin{equation}
	\psi_\sigma(x) = \frac{1}{\sqrt{2\pi a_0}}F_{\sigma}e^{-i\sigma 2\pi N_\sigma x/L}e^{-i\phi_\sigma(x)},
\end{equation}
where $F_\sigma$ is a Klein factor and
\begin{equation}
	\phi_\sigma(x)= -\sum_{n_q=1}^\infty \frac{1}{\sqrt{n_q}} e^{- \pi n_q a_{0}/L}\left( e^{-iqx}b_{q\sigma} + e^{iqx}b^\dagger_{q\sigma} \right).
\end{equation}

The correlator then takes the form
\begin{widetext}
\begin{align}
	R_x(x)={}&-\tfrac{1}{2}\operatorname{Re}
	\frac{1}{(2\pi a_{0})^2} \langle e^{i\phi_\uparrow(x)}e^{i 2\pi N_{\uparrow}x/L}F^\dagger_{\uparrow}F^{\vphantom{\dagger}}_{\uparrow}e^{-i\phi_\uparrow(0)}e^{i\phi_\downarrow(0)}F^\dagger_{\downarrow}F^{\vphantom{\dagger}}_{\downarrow}e^{i 2\pi N_\downarrow x/L} e^{-i\phi_\downarrow(x)}\rangle\nonumber\\
	=-{}&\tfrac{1}{2}\operatorname{Re}\frac{1}{(2\pi a_{0})^2} \langle e^{i 2\pi (N_\downarrow+ N_\uparrow) x/L} \rangle \langle e^{i\phi_\uparrow(x)}e^{-i\phi_\uparrow(0)}e^{i\phi_\downarrow(0)} e^{-i\phi_\downarrow(x)}\rangle, \label{rxboson}
\end{align}
where we used the identity $F^\dagger_{\sigma}F^{\vphantom{\dagger}}_{\sigma'} = \delta_{\sigma\sigma'}$. In the second equality the expectation value has been factored into a product of two expectation values, which is allowed because every state can be completely and uniquely described by the number of fermions $ N_{\sigma}$ and bosonic excitations $b^\dagger_{q\sigma }b^{\vphantom{\dagger}}_{q\sigma }$, implying the independence of their expectation values.

The next step is to transform to the eigenbasis of the $B$-operators,
\begin{equation}
	\Phi_\sigma(x)= -\sum_{n_q=1}^\infty \frac{1}{\sqrt{n_q}} e^{- \pi n_q a_{0}/L}\left( e^{-iqx}B_{q\sigma} + e^{iqx}B^\dagger_{q\sigma} \right),
\end{equation}
by means of the relations
\begin{subequations}
\begin{align}
	\phi_{\uparrow} ={}&- 2^{-3/2}( K^{-1/2}+ K^{1/2})[ \Phi_{+}(x)+\Phi_{-}(-x) ] - 2^{-3/2}( K^{-1/2}- K^{1/2})[ \Phi_{+}(-x)-\Phi_{-}(x) ] ,\\
	\phi_{\downarrow} ={}&- 2^{-3/2}( K^{-1/2}+ K^{1/2})[ \Phi_{-}(x)-\Phi_{+}(-x) ] + 2^{-3/2}( K^{-1/2}- K^{1/2})[\Phi_{-}(-x)+\Phi_{+}(x) ] .
\end{align}
\end{subequations}
We thus evaluate the expectation value
\begin{align}
	\langle e^{i\phi_\uparrow(x)}e^{-i\phi_\uparrow(0)}e^{i\phi_\downarrow(0)} e^{-i\phi_\downarrow(x)}\rangle	={}&\big\langle e^{-i\sqrt{\frac{K}{2}}\left( \Phi_{+}(x) + \Phi_{+}(-x)-2\Phi_+(0) \right) } \big\rangle \big\langle e^{-i\sqrt{\frac{K}{2}} \left( \Phi_{-}(x) - \Phi_{-}(-x) \right) }\big\rangle\nonumber\\
	&\times  e^{[\Phi_{+}(x), \Phi_{+}(0)]}e^{-\frac{1}{4} [\Phi_{+}(x), \Phi_{+}(-x)]} e^{-\frac{1}{4} [\Phi_{-}(x), \Phi_{-}(-x)]}, \label{bcorr_eig}
\end{align}
where we used the Baker-Campbell-Hausdorff formula and the 
fact that the commutator $[\Phi_{\sigma}(x), \Phi_{\sigma}(y)]$ is a $c$-number. 

Because of inversion symmetry,
\begin{equation}
[\Phi_{+}(x),\Phi_{+}(-x)]=[\Phi_{-}(-x),\Phi_{-}(x)]=-[\Phi_{-}(x),\Phi_{-}(-x)],
\end{equation}
the last two exponentials in Eq.\ \eqref{bcorr_eig} cancel each other. The identity for the thermal average $ \langle e^{B} \rangle= e^{\langle  B^2\rangle/2}$ of an operator $B$ that is linear in free bosonic operators then gives
\begin{subequations}
\begin{align}
	&\big\langle e^{-i\sqrt{\frac{K}{2}}\left( \Phi_{+}(x) + \Phi_{+}(-x)-2\Phi_+(0) \right) } \big\rangle	= e^{-\frac{K}{4}\langle 6\Phi_+^2(0) + \{\Phi_+(x),\Phi_+(-x)\} - 4\{\Phi_+(x),\Phi_+(0)\} \rangle },\\
	&\big\langle e^{-i\sqrt{\frac{K}{2}}\left( \Phi_{-}(x) - \Phi_{-}(-x) \right) } \big\rangle	= e^{-\frac{K}{4}\langle 2\Phi_-^2(0) - \{\Phi_-(x),\Phi_-(-x)\} \rangle },
\end{align}
\end{subequations}
where we also used the translational symmetry $\langle \Phi_{+}(0)\Phi_{+}(x) \rangle = \langle \Phi_{+}(-x)\Phi_{+}(0) \rangle $. We need one more identity,
\begin{equation}
  e^{\langle\{\Phi_-(x),\Phi_-(-x)\}\rangle - \langle\{\Phi_+(x),\Phi_+(-x)\}\rangle} = 1,
\end{equation}
again because of inversion symmetry, to finally arrive at
\begin{equation}
	R_x(x)= -\tfrac{1}{2}\operatorname{Re}\frac{1}{(2\pi a_{0})^2}\langle e^{i 2\pi N_{\rm tot} x/L} \rangle e^{K\langle \{\Phi_+(x),\Phi_+(0)\}-2\Phi_+^2(0)  \rangle } e^{[\Phi_{+}(x), \Phi_{+}(0)]},
\end{equation} 
with $N_{\rm tot}=N_\uparrow+N_\downarrow$. 

\subsection{Evaluation of the thermal averages}

It remains to thermally average the bosonic field correlators and the exponential of the fermionic number operators. We do the latter average first.

The number operators $ N_\uparrow,N_\downarrow$ commute with each other and with the bosonic fields, so the average is a classical ensemble average with the Gibbs measure at inverse temperature $\beta$ and chemical potential $\mu$,  
\begin{subequations}
\label{FNdef}
\begin{align}
&G_N(x)\equiv \langle e^{i 2\pi N_{\rm tot} x/L} \rangle
=\frac{\sum_{N_\uparrow,N_\downarrow=-\infty}^\infty e^{i 2\pi N_{\rm tot} x/L} e^{-\beta E+\beta\mu N_{\rm tot}}}{\sum_{N_\uparrow,N_\downarrow=-\infty}^\infty e^{-\beta E+\beta\mu N_{\rm tot}}},\\
&E=\frac{2\pi v_{\rm F}}{L}\biggr[\tfrac{1}{2}  N_\uparrow(  N_\uparrow + 1)+\tfrac{1}{2}  N_\downarrow(  N_\downarrow + 1)  + \kappa  N_{\uparrow} N_{\downarrow}\biggr].
\end{align}
\end{subequations}

The chemical potential $\mu=0$ without interactions, corresponding to a half-filled band at zero temperature. To keep the half-filled band also for nonzero $\kappa$ we adjust
\begin{equation}
\mu=-\tfrac{1}{2}\kappa \frac{2\pi v_{\rm F}}{L}.
\end{equation}
Then $\lim_{\beta\rightarrow \infty}\langle N_\sigma\rangle =-1/2$ independent of the interaction strength $\kappa\in(-1,1)$, as required by Eq.\ \eqref{Nonehalf}.

We next turn to the average of the bosonic field correlators:
\begin{align}
	\langle \Phi_{\sigma}(x)\Phi_{\sigma}(0)\rangle={}& \sum_{n_q, n_q'=1}^\infty\frac{1}{\sqrt{n_q n_q'}}e^{- 2\pi n_q a_{0}/L}\langle\bigl(e^{-iqx}B_{q\sigma}+ e^{iqx}B^\dagger_{q\sigma}\bigr)\bigl(B_{q'\sigma}+B^\dagger_{q'\sigma}\bigr)\rangle\nonumber\\
	={}&\sum_{n_q=1}^\infty\frac{1}{n_q }e^{-2 \pi n_q a_{0}/L}\langle e^{-iqx}B_{q\sigma}B^\dagger_{q\sigma} + e^{iqx}B^\dagger_{q\sigma}B_{q\sigma}\rangle\nonumber\\
	={}&\sum_{n_q=1}^\infty\frac{1}{n_q }e^{-2 \pi n_q a_{0}/L}\left(\frac{e^{-i\sigma 2\pi n_qx/L}}{1-e^{- n_q 2\pi v\beta/L}} + \frac{e^{i\sigma 2\pi n_qx/L}}{e^{  n_q 2\pi v\beta/L}-1}\right) \nonumber\\
	={}&\sum_{n_q=1}^\infty\frac{1}{n_q}e^{-n_q(2\pi/L)(a_0+i\sigma x)}+\sum_{n_q=1}^\infty\frac{1}{n_q}e^{-2\pi  n_q a_0/L }\frac{2\cos(2\pi n_q x/L)}{e^{  n_q 2\pi v\beta/L}-1} \nonumber\\
	={}& - \ln\left(1-e^{-(2\pi/L)(a_0+i\sigma x)}\right)+ \sum_{n_q=1}^\infty\frac{1}{n_q}e^{-2\pi  n_q a_0/L }\frac{2\cos(2\pi n_q x/L)}{e^{  n_q 2\pi v\beta/L}-1}.\label{bf_corr}
\end{align}
We subtract $\langle\Phi^2_{\sigma}(0)\rangle$ and expand to first order in the cutoff length $a_0$,
\begin{equation}
\begin{split}
	&\langle \Phi_{\sigma}(x)\Phi_{\sigma}(0) - \Phi^2_{\sigma}(0)\rangle = G_\sigma(x)+\ln(2\pi a_0/L)+{\cal O}(a_0),\\
	&G_\sigma(x)= - \ln(1-e^{-i\sigma  2\pi x/L})+  \sum_{n_q=1}^\infty\frac{4}{n_q}\frac{\sin^2(\pi  n_q x/L)}{e^{  n_q 2\pi v\beta/L}-1}.
	\end{split}\label{bf_corr2}
\end{equation}

With the help of two further identities,
\begin{subequations}
\begin{align}
&\langle \{\Phi_+(x),\Phi_+(0)\}-2\Phi_+^2(0)  \rangle = 2\operatorname{Re}\langle \Phi_{+}(x)\Phi_{+}(0) - \Phi^2_{+}(0)\rangle,\\
&[\Phi_{+}(x), \Phi_{+}(0)] = 2i\operatorname{Im}\langle \Phi_{+}(x)\Phi_{+}(0) - \Phi^2_{+}(0)\rangle=2\pi i(x/L-1/2),
\end{align}
\end{subequations}
\end{widetext}	
we conclude that
\begin{equation}
	R_x(x)
	= \frac{ e^{2K \operatorname{Re}G_+(x) }\operatorname{Re}\bigl[ e^{i2\pi x/L}G_N(x)  \bigr]}{2L^2(2\pi a_0/L)^{2-2K}}.
\end{equation}
The two functions $G_N(x)$ and $G_+(x)$ can be computed efficiently from Eqs.\ \eqref{FNdef} and \eqref{bf_corr2}. For the comparison with the lattice theory we identify the cutoff length $a_0$ with the lattice constant $a$.

\subsection{Propagator}

For reference we also give the finite-size bosonization result for the propagator:
\begin{align}
	C_\sigma(x)={}&\langle\psi_\sigma^\dagger(x,0)\psi_\sigma^{\vphantom{\dagger}}(0,0)\rangle\nonumber\\
	={}&\frac{1}{2\pi a_{0}} \langle e^{i \sigma 2\pi N_\sigma x/L} \rangle \langle e^{i\phi_\sigma(x)}e^{-i\phi_\sigma(0)}\rangle\nonumber\\
	={}&\frac{\sigma}{2\pi i a_0}\left(\frac{2\pi a_0}{L}\right)^{(1/2)(K+1/K)}\langle e^{i \sigma\pi (2N_\sigma+1) x/L} \rangle\nonumber\\
	&\times e^{(1/2)(K+1/K)\operatorname{Re}G_\sigma(x)}. \label{Csigmaboson}
\end{align}


\begin{thebibliography}{99}
\bibitem{note1} We set the branch cut of the logarithm along the negative real axis, so $\ln e^{ika}$ is a sawtooth that jumps at $ka=\pi+2n\pi$.
\bibitem{note2} The continuum limit $C_\sigma(k)=-\tfrac{1}{2}\operatorname{sign}(\sigma k)$ differs from the zero-temperature Fermi function $\theta(-\sigma k)$ by a $1/2$ offset. This offset corresponds to a delta function $\delta(x-x')$ contribution to the propagator \eqref{Csigmaxxdef}, which is lost in the discretization.
\end{thebibliography}
\end{document}